\documentclass[a4paper,10pt]{article}
\usepackage{amsmath}
\usepackage{float}
\usepackage{bm}
\usepackage{comment}
\usepackage{booktabs}
\usepackage{subfigure}
\usepackage{cite}
\usepackage{authblk}
\usepackage{array}

\usepackage{color}

\newcommand{\green}[1]{}

\usepackage{graphicx}
\title{Iterated crowdsourcing dilemma game}
\author[1,2]{Koji Oishi}
\author[3]{Manuel Cebrian}
\author[3]{Andres Abeliuk}
\author[4,2]{Naoki Masuda\thanks{Corresponding author: masuda@mist.i.u-tokyo.ac.jp}}
\affil[1]{Department of Applied Physics, Graduate School of Engineering, The University of Tokyo, 7-3-1 Hongo, Bunkyo, Tokyo 113-8656, Japan}
\affil[2]{CREST, JST, 4-1-8 Honcho, Kawaguchi, Saitama 332-0012, Japan.}
\affil[3]{National Information and Communications Technology Australia, University of Melbourne, Victoria 3010, Australia}
\affil[4]{Department of Mathematical Informatics, The University of Tokyo, 7-3-1 Hongo, Bunkyo, Tokyo 113-8656, Japan}

\begin{document}
\maketitle


The Internet has enabled the emergence of collective problem solving, also known as crowdsourcing, as a viable option for solving complex tasks. However, the openness of crowdsourcing presents a challenge because solutions obtained by it can be sabotaged, stolen, and manipulated at a low cost for the attacker. We extend a previously proposed crowdsourcing dilemma game to an iterated game to address this question. We enumerate pure evolutionarily stable strategies within the class of so-called reactive strategies, i.e., those depending on the last action of the opponent. Among the 4096 possible reactive strategies, we find 16 strategies each of which is stable in some parameter regions. Repeated encounters of the players can improve social welfare when the damage inflicted by an attack and the cost of attack are both small. Under the current framework, repeated interactions do not really ameliorate the crowdsourcing dilemma in a majority of the parameter space.

\newpage


Crowdsourcing has opened a plethora of possibilities for individuals around the world to connect, coordinate, and solve complex problems that are currently beyond computational capabilities~\cite{howe2006rise,von2006games,von2008recaptcha,huberman2009crowdsourcing,cooper2010predicting,hand2010citizen,horowitz2010anatomy,hellerstein2011searching,pickard2011time,tang2011reflecting,barrington2012game,mason2012conducting,zhang2012task,alstott2013predictors,rahwan2013global}. At the same time, a number of problems have arisen by the use of this novel technology. In particular, the openness of crowdsourcing presents individuals with an opportunity to exhibit antisocial behavior such as plagiarizing, sabotaging, and manipulating the solution being collectively obtained (see Refs.~\cite{Manuel} and~\cite{cebriannas} for brief reviews).

Although techniques for securing crowdsourcing operations have been expanding steadily, so has the number of applications of crowdsourcing~\cite{FutureCrowd}. As a result, a silver bullet to secure crowdsourcing for all possible attacks may be difficult to find. Services such as Amazon's Mechanical Turk will likely diminish the problem of intentional attacks by using a reputation system, discouraging participants to sabotage~\cite{mason2012conducting}. Other approaches such as error correction have also been shown to be effective in crowdsourcing settings. However, they are limited in their applicability to specific contexts~\cite{Iperiotis}. In this study, we consider the
possibility that repeated encounters between the same peers
alleviate sabotage.

Motivated in part by the DARPA Network
Challenge~\cite{pickard2011time,tang2011reflecting}, a crowdsourcing dilemma game
in which two competing firms interact in a two-stage game was recently
proposed~\cite{Manuel}. In the
first stage, each of the two firms selects whether or not to achieve a
given task via crowdsourcing. If the firm decides not to crowdsource,
it tries to solve the problem in-house. In the second stage, the firms
have the option of attacking the opponent if the opponent has selected
to solve the task via crowdsourcing. The equilibrium strategies of the model
depend on complex tradeoffs between the productivity value, the
benefit of attack, and the cost of attack. In summary, there are three
parameter regions. First, crowdsourcing by both agents is the unique
equilibrium when the damage inflicted by an attack is low. Second, the
in-house solution (i.e., not crowdsourcing) selected by both agents is
the unique equilibrium when the damage inflicted by an attack is high
and the cost of attack is low. Third, the crowdsourcing by both agents
and the in-house solution of both agents are two equilibria when the
damage inflicted by an attack is high and the cost of attack is
high. 

In the crowdsourcing dilemma game \cite{Manuel}, attacking the opponent's task that has been crowdsourced lessens the welfare of both parties, which 
is a social dilemma. In the theory of cooperation
in social dilemma situations, there have been proposed various mechanisms 
to evade socially undesirable equilibria.
One such mechanism is iterated interaction, also called direct reciprocity,
which has been successful in realizing mutual cooperation in the prisoner's dilemma game~\cite{Trivers-1971,Axelrod-book-1984}.
It may be possible to alleviate the crowdsourcing dilemma by similar repeated
encounters between players in crowdsourcing competitions such that
a socially desirable state such as
a decreased level of attacks emerges.

In fact, mutual cooperation emerges in the iterated prisoner's dilemma
under appropriate conditions if players adopt conditional strategies
such as variants of Tit-for-Tat (i.e., do what the opponent did in the last round) \cite{Trivers-1971,Axelrod-book-1984,Nowak-Sigmund-1992,Kraines-Kraines-1993,Nowak-Sigmund-1993,Nowak-book-2006}.
In the present study, we examine a full range of conditional
strategies by formulating a variant of the crowdsourcing dilemma game
as an iterated game. For a computational reason, we restrict ourselves
to the strategies that use the information about
the action of the opponent in the previous encounter.

\section*{Results}

\subsection*{Model}

To examine evolutionarily stable strategies (ESSs), we consider an infinitely large well-mixed population of players in which two randomly selected players are engaged in an iterated crowdsourcing dilemma game.
Each player is engaged in the game sufficiently many times in one generation.

\begin{figure}[!htb]
\begin{center}
\includegraphics[width=140mm]{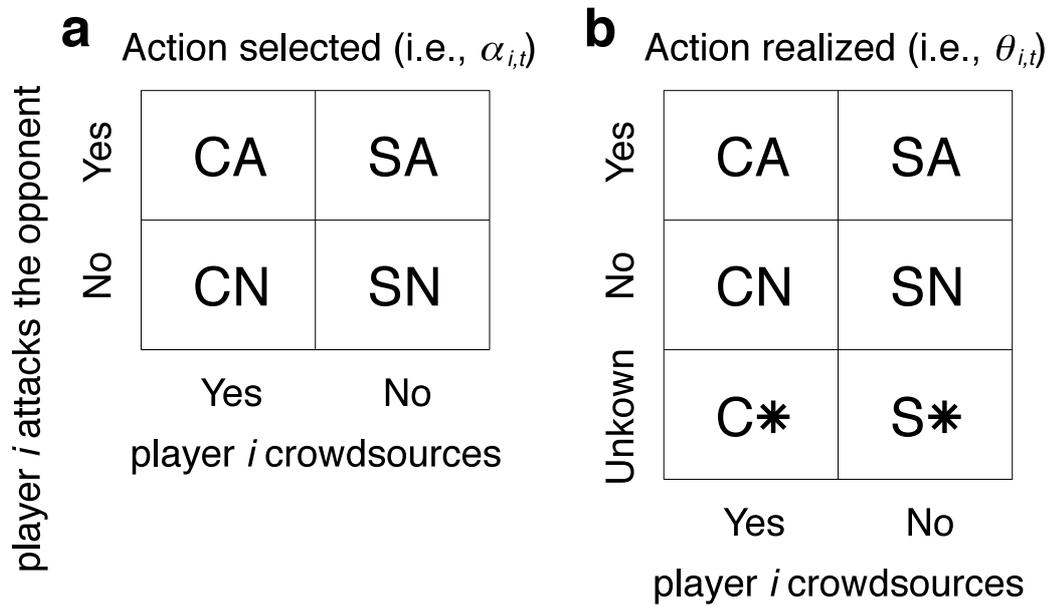}
\caption{Actions selected and realized in a single round. (a) Four types of actions selected by a player in a single round. (b) Six types of actions realized by a player in a single round.}
\label{fig:strategy}
\end{center}
\end{figure}

Consider the iterated game between players 1 and 2.
In every round of the game, each player submits an action, which generally depends on the action of the opponent in the previous round.
We denote the action selected by player $i \in \{1,2\}$ in round $t$ by $\alpha_{i,t}$, which is either CA, CN, SA, or SN (Figure~\ref{fig:strategy}(a)).
With $\alpha_{i,t}=\text{CA}$ ($\alpha_{i,t}=\text{SA}$), player $i$ selects to crowdsource (not to crowdsource) and attack the opponent if the opponent crowdsources in round $t$. With
$\alpha_{i,t}=\text{CN}$ ($\alpha_{i,t}=\text{SN}$), player $i$ selects to crowdsource (not to crowdsource) and not to attack the opponent if the opponent crowdsources in round $t$.
It should be noted that CA and CN are behaviorally the same unless the opponent crowdsources.
In this case, the opponent that has not crowdsourced does not know whether the focal player has selected CA or CN.
By the same token, SA and SN are the same unless the opponent crowdsources.

There are six types of action that a player $i$ realizes in a single round
(Figure~\ref{fig:strategy}(b)).
We denote the realized action of player $i$ in round $t$ by $\theta_{i,t}$, which is either CA, CN, C{\textasteriskcentered}, SA, SN, or S{\textasteriskcentered}.
$\theta_{i,t}=\text{CA}$ ($\theta_{i,t}=\text{SA}$) means that player $i$ has crowdsourced (has not crowdsourced) and attacked the opponent.
$\theta_{i,t}=\text{CN}$ ($\theta_{i,t}=\text{SN}$) means that player $i$ has crowdsourced (has not crowdsourced) and has not attacked the opponent.
$\theta_{i,t}=\text{C{\textasteriskcentered}}$ ($\theta_{i,t}=\text{S{\textasteriskcentered}}$) means that player $i$ has crowdsourced (has not crowdsourced) and that whether player $i$ has intended to attack the opponent or not is unknown to the opponent.
If $\theta_{i,t}$ is either CA, CN, SA, or SN, $i$'s opponent has crowdsourced in round $t$.
If $\theta_{i,t}$ is either C{\textasteriskcentered} or S{\textasteriskcentered}, $i$'s opponent has not crowdsourced.
The relationship between the actions selected by the two players and the realized actions perceived by the two players is shown in Table~\ref{submit_and_realize}.

We consider players adopting the so-called reactive strategies \cite{Nowak-Sigmund-1992,Kraines-Kraines-1993,Nowak-book-2006,Nowak-Sigmund-1990}.
A player adopting a reactive strategy selects an action based on the opponent's realized action in the previous round.
Therefore, a reactive strategy of player 1 is a mapping from $\theta_{2,t}$ to $\alpha_{1,t+1}$.
There are $4^{6}=4096$ reactive strategies.

We assume that players commit an action implementation error with a small probability $\epsilon$ ($0<\epsilon \ll 1$).
For simplicity, the decision of crowdsourcing and that of attacking are assumed to err independently with the same probability $\epsilon$.
For example, player intending CA actually carries out CA with probability $(1-\epsilon)^2$, CN with probability $\epsilon(1-\epsilon)$, SA with probability $\epsilon(1-\epsilon)$, and SN with probability $\epsilon^2$.

The payoff in a round is determined in the same way as in the original
crowdsourcing dilemma game \cite{Manuel}.  A player's productivity value is equal to zero as normalization
when the player does not crowdsource. It obeys the uniform distribution on $(0,1)$ when the player crowdsources.  A
player needs to pay cost $q \in (0,1)$ to attack the opponent to
reduce the opponent's productivity by $d \in (0,1)$.  The player that
finally obtains the higher productivity than the opponent wins the
unitary payoff in the current round. The other player gains nothing.
If the productivity values of the two players are the same,
each player wins with probability $1/2$.  
It should be noted that players decide the actions without referring to
the productivity values of the player itself and the opponent.

We do not consider time discounting of the payoff across rounds and do assume that the number of rounds is very large. Therefore, we are concerned with the stationary state of the actions adopted by the two players and the payoff per round.

Even if we confine ourselves to a single-round game, the present model
is slightly different from the previous model \cite{Manuel} in the following aspects. First, in the previous model \cite{Manuel}, it was assumed that the productivity values of both players thanks to crowdsourcing were unknown to each player when the players determined whether to crowdsource or not in the first stage. The productivity values were then revealed just before the second stage occurred. In other words, if the opponent has crowdsourced, the focal player knows the opponent's productivity (and the focal player's own productivity if the focal player has crowdsourced) before they determine whether to attack the opponent or not. Therefore, each player is assumed to obey the best response rule in the second stage. In contrast, in the present model, we assumed that the players select the actions for the first 
stage (i.e., crowdsource or not to crowdsource) 
and the second stage (i.e., attack or not to attack) in the beginning of the round without knowing the productivity of the players in the middle of the round. We changed the model in this way because, otherwise, there are a continuum of pure strategies because of the productivity is continuously valued. By confining ourselves to a model with a finite set of discrete pure strategies, we aim to carry out an exhaustive and rigorous analysis of the model to understand the iterated as well as non-iterated crowdsourcing dilemma game.

\subsection*{Non-iterated game}

\begin{figure}[!htb]
\begin{center}
\includegraphics[width=87mm]{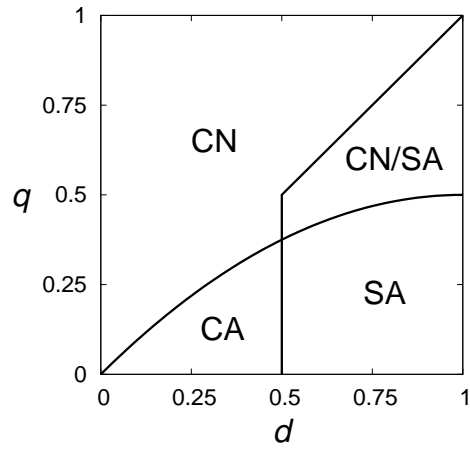}
\caption{Phase diagram for the single-shot game. In the region labeled CN/SA, both CN and SA are ESSs.}
\label{fig:single-shot phase diagram}
\end{center}
\end{figure}

We started by analyzing the non-iterated crowdsourcing dilemma game. Because strategies conditioned on the realized action in the previous round are irrelevant, there are four pure strategies, i.e., CA, CN, SA, and SN.
The ESSs in the full $(d,q)$ parameter space are shown in Figure~\ref{fig:single-shot phase diagram}. The figure indicates that crowdsourcing is stable when the damage inflicted by an attack (i.e., $d$) is small or the cost of attack (i.e., $q$) is large. Attacking is stable when $d$ is large or $q$ is small.

The results shown in Figure~\ref{fig:single-shot phase diagram}
are qualitatively the same as those for the previously analyzed single-shot crowdsourcing dilemma game \cite{Manuel} in the meaning that crowdsourcing is stable when $d$ is small or $q$ is large. In contrast, a large $q$ value does not prevent the players from attacking the opponent in the previous model \cite{Manuel}, whereas not to attack is an ESS for large $q$ (irrespectively of $d$) in the present model.

\subsection*{Evolutionary stability and efficiency for the iterated game}

\begin{figure}[!htb]
\begin{center}
\includegraphics[width=140mm]{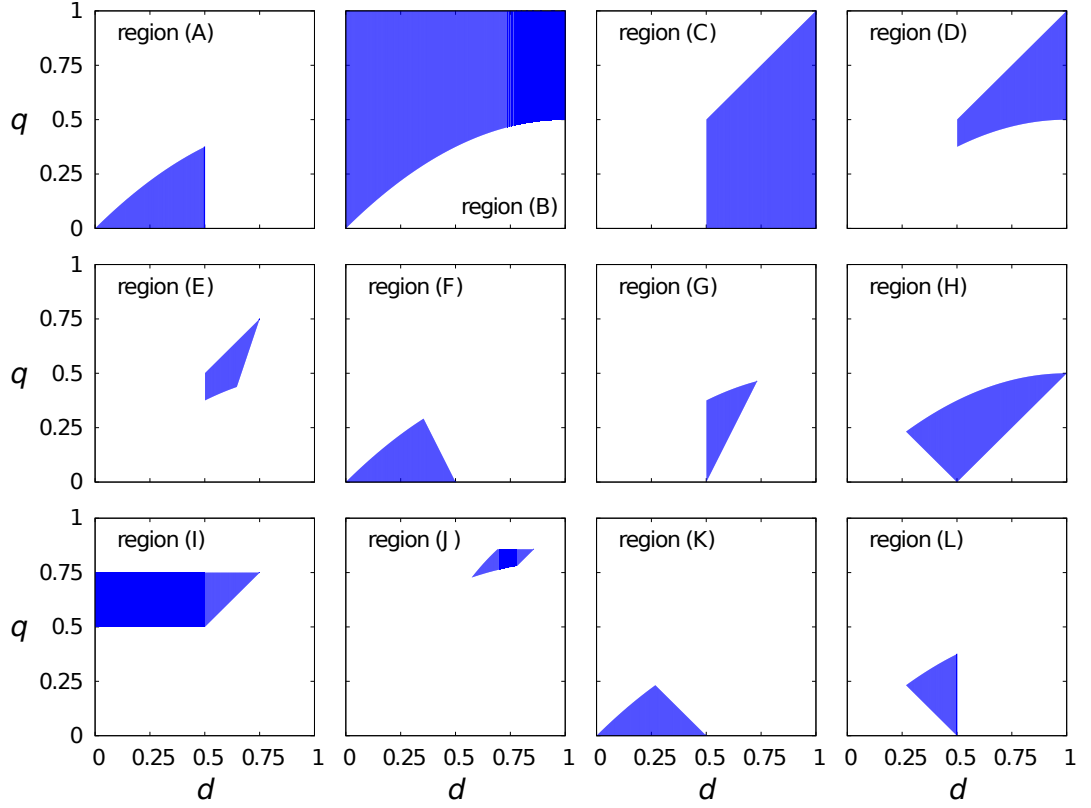}
\caption{Parameter regions (A) to (L) in the $(d,q)$ space.
In addition to the trivial condition $0<d, q<1$, the 12 regions are defined as follows. Region (A): $d<1/2$ and $q<(1/2)d(2-d)$. Region (B): $q>(1/2)d(2-d)$. Region (C): $d>1/2$ and $q<d$. Region (D): $d>1/2$ and $(1/2)d(2-d)<q<d$. Region (E): $d>1/2$ and $\text{max}\{(1/2)d(2-d),(3/2)(2d-1)\}<q<d$.
Region (F): $q<\text{min}\{(1/2)d(2-d),1-2d\}$. Region (G): $d>1/2$ and $2d-1<q<(1/2)d(2-d)$. Region (H): $\text{max}\{1/2-d,d-1/2\}<q<(1/2)d(2-d)$. Region (I): $\text{max}\{d,1/2\}<q<3/4$. Region (J): $\text{max}\{(2/5)(1+2d-d^2),d\}<q<\text{min}\{(1/2)(-1+6d-3d^2),6/7\}$. Region (K): $q<\text{min}\{(1/2)d(2-d),1/2-d\}$. Region (L): $d<1/2$ and $1/2-d<q<(1/2)d(2-d)$.}
\label{region_plot}
\end{center}
\end{figure}

We exhaustively searched ESSs among the 4096 reactive strategies.
We found 16 strategies that were ESSs in some regions of the $(d,q)$ parameter space.
The 16 ESSs are listed in Table~\ref{ess_def_region}.
In the table, $\alpha_n(\text{SN})$, for example, indicates the action selected when the opponent realized SN in the previous round.
Each strategy is an ESS in the parameter region specified by the label (one of (A) through (J)) shown in the table.
The parameter regions are depicted in Figure~\ref{region_plot} (see the caption for the precise definition).

Ten out of the 16 ESSs are efficient for some $d$ and $q$ values.
The parameter regions in which these ESSs are efficient are shown in Table~\ref{ess_def_region}.
We checked the condition for the efficiency by referring to the average payoffs of ESSs in the homogeneous population (Table~\ref{tab:stationary}).

Strategies 1, 2, and 3 are unconditional strategies, whereas the other strategies are conditional strategies.
We call the three unconditional strategies uncond-CA, uncond-CN, and uncond-SA, respectively.

In parameter regions (B) and (C), uncond-CN and uncond-SA are efficient ESSs, respectively. This result is the same as that for the single-shot game (Figure~\ref{fig:single-shot phase diagram}).
In the intersection of regions (B) and (C), which is region (D), strategies 4, 5, 6, 7, 8, and 9 are also efficient ESSs.
In subregions of (B) and (C), strategies 10, 11, 13, 15, and 16 are inefficient ESSs.
Strategies 10, 11, and 15, but not 13 yield the same payoff as the efficient ESSs in the limit $\epsilon\to 0$.
Strategy 13 is the only ESS that yields a smaller payoff than that of the coexisting unconditional ESS (i.e., uncond-SA) in the limit $\epsilon\to 0$.

In region (A), neither uncond-CN nor uncond-SA is an ESS, and uncond-CA is an inefficient ESS. 
Instead, strategy 12 or 14, both of which are conditional strategies, is the efficient ESS in the region.
It should be noted that, in the single-shot game, (uncond-)CA is 
the unique ESS in this parameter region (Figure~\ref{fig:single-shot phase diagram}). Because regions (A), (B), and (C) exhaust the entire parameter space $0<d<1$, $0<q<1$, conditional ESSs yield larger payoffs than unconditional ESSs only in region (A).
In other words, making the crowdsourcing dilemma game an iterated game improves the efficiency of the ESS exclusively in this parameter region.

Region (A) is composed of subregions (K) and (L).

In region (K), strategy 14 is not an ESS, and strategy 12 is the efficient ESS.
The payoff of strategy 12 in the homogeneous population is larger than that of uncond-CA by $(1/2)q\epsilon+\mathcal{O}(\epsilon^2)$.
The difference vanishes in the limit $\epsilon\to 0$.
This is because a pair of players adopting strategy 12 almost always implements CA for infinitesimally small $\epsilon$.

In spite of this similarity between strategy 12 and uncond-CA, strategy 12 is efficient because, when a pair of players adopts strategy 12, their realized actions persist in SA for some time once both players start implementing SA.
To understand this phenomenon, consider the situation in which both players adopting strategy 12 implement CA. This situation almost always occurs in the limit $\epsilon\to 0$.
The two players simultaneously switch to SA if both players commit an error to select either SA or SN in the same round.
This event occurs with probability $\left[\epsilon\left(1-\epsilon\right)+\epsilon^2\right]^2=\epsilon^2$.
They return to selecting CA if either player commits an error to select CA or CN. This event occurs with probability $1-\left[1-\epsilon\left(1-\epsilon\right)-\epsilon^2\right]^2=2\epsilon-\epsilon^2$.
Therefore, the fraction of the number of rounds in which the two players implement SA is approximately equal to $(1/2)\epsilon$ for small $\epsilon$.
During the period in which the two players implement SA, the cost of attack is evaded.
In contrast, a player adopting uncond-CA cannot avoid the cost of attack, irrespective of whether the opponent adopts uncond-CA or strategy 12.
This is because the repetition of SA does not persist and both players almost always implement CA.
Therefore, strategy 12 is stable against invasion by uncond-CA and yields a slightly larger payoff than uncond-CA in the homogeneous population.

In region (L), strategy 14 is the efficient ESS.
The payoff of strategy 14 in the homogeneous population is larger than that of uncond-CA by $q+\mathcal{O}(\epsilon)$, which does not vanish for infinitesimally small $\epsilon$.

A pair of players adopting uncond-CA almost always implements CA and obtains $1/2-q$ per round for infinitesimally small $\epsilon$. This is because both players pay the cost of attack (i.e., $q$) and win the game with probability $1/2$.
In contrast, a pair of players adopting strategy 14 almost always implements SA for infinitesimally small $\epsilon$, as shown in Table~\ref{tab:stationary}.
The two players obtain $1/2$ per round; each player wins with probability $1/2$ without paying the cost of attack.
This is the reason why strategy 14 yield a larger payoff than uncond-CA in the limit $\epsilon\to 0$.
A player adopting strategy 14 alternates between CA and SA in the absence of error if the opponent adopts uncond-CA.
In this situation, the average payoff of the opponent is equal to $(1/2)\left\{(1-d)+(1/2-q)\right\}$.
Therefore, strategy 14 is stable against invasion by uncond-CA when $q>1/2-d$.
This condition defines a boundary of region (L).
It should be noted that, when $d<1/2$, i.e., when being attacked is not so costly, players gain a larger payoff by selecting CA rather than SA irrespective of the action of the opponent.
Therefore, in contrast to strategy 14, uncond-SA is not stable in region (L).

\subsection*{Size of attractive basins of different ESSs in parameter region (A)}

\begin{figure}[!htb]
\begin{center}
\includegraphics[width=140mm]{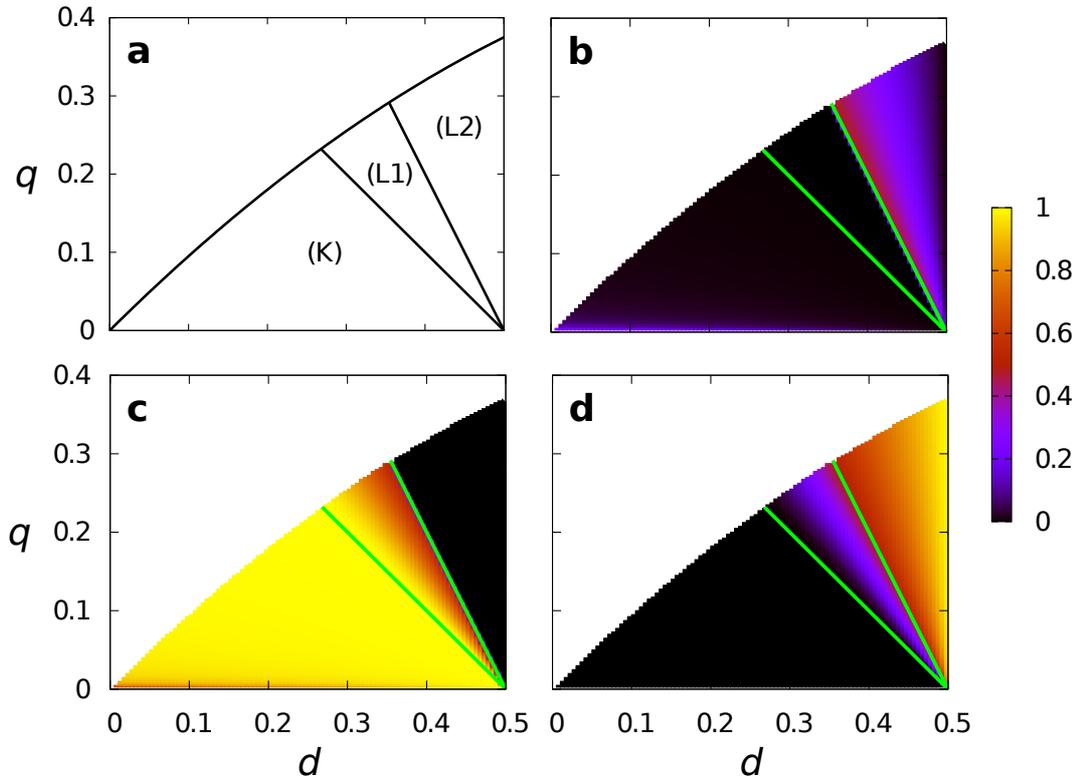}
\caption{Details of region (A). (a) Three subregions of region (A) in the $(d,q)$ space. (b), (c), (d) Relative sizes of the attractive basins. (b) uncond-CA, (c) strategy 12, and (d) strategy 14. We set $\epsilon=10^{-3}$. The two solid lines in (b), (c), and (d) represent the boundaries between the subregions. The size of attractive basin of strategies 12 and 14 is equal to zero in region (L2) and region (K), respectively, because they are not ESSs in the corresponding region.}
\label{basin}
\end{center}
\end{figure}

In the previous section, we revealed that conditional strategies
were the only efficient ESSs in region (A). For these conditional
strategies to establish a foothold in an evolutionary context, they
should also have a sufficiently large attractive basin under
evolutionary dynamics.  Therefore, we compare the relative size of the
attractive basins of the ESSs in region (A).  We examine replicator
dynamics composed only of ESSs because it is not feasible to treat the
dynamics composed of all 4096 strategies.  Region (K) allows two ESSs,
i.e., uncond-CA and strategy 12.  Region (L) is divided into
subregion (L1) that allows three ESSs, i.e., uncond-CA, strategy 12,
and strategy 14, and subregion (L2) that allows two ESSs, i.e.,
uncond-CA and strategy 14. The boundary between (L1) and (L2) is
given by $q=1-2d$. These regions within region (A) are depicted in Figure~\ref{basin}.  We separately calculated the size of attractive
basins for regions (K), (L1), and (L2).

The size of the attractive basin for each ESS is shown in Figure~\ref{basin} with $\epsilon=10^{-3}$.
The attractive basins of strategies 12 and 14 are larger than that of uncond-CA for a large parameter region in region (A).
In particular, strategies 12 and 14 have the largest attractive basin in most of region (K) and the entire region (L2), respectively. Therefore, conditional strategies 12 and 14 are not only efficient but also reached from various initial conditions under replicator dynamics.

\section*{Discussion}

We explored the possibility of improving the quality of solution in a crowdsourcing game by analyzing an iterated game. We found that the crowdsourcing dilemma was alleviated when the damage inflicted by an attack (i.e., $d$) and the cost of attack (i.e., $q$) were small (i.e., region (A)). In this parameter region, 
an unconditional strategy (uncond-CA) and 
either conditional strategy (strategy 12 or 14) are coexisting ESSs. Furthermore, repetition of the game allows the emergence of the equilibria (i.e., strategies 12 and 14) that are more efficient than the equilibrium for the single-shot game (i.e., uncond-CA). In the other parameter regions, repeated encounters do not alter the efficient ESSs relative to the case of the non-iterated game.

Strategy uncond-CA is analogous to unconditional defection in the
prisoner's dilemma game. Strategies 12 and 14 are analogous to
retaliative strategies in the prisoner's dilemma game. However, we
emphasize that, the loose analogue between the crowdsourcing dilemma
game and the prisoner's dilemma game is only justified in region (A).
Because strategy 12 is only marginally superior
to uncond-CA in region (A), we discuss the combat between strategy 14
and uncond-CA; the homogeneous population of strategy 14 yields the
payoff that is larger by $\approx q$ than that realized by the
homogeneous population of uncond-CA. If strategy 14 and uncond-CA play the iterated game, each
strategist almost always selects CA or SA. CA and SA are analogous to
defection and cooperation in the prisoner's dilemma, respectively. If
both selects SA, both players obtain $1/2$ in a single round
if $\mathcal{O}(\epsilon)$ terms are neglected. If the focal player
selects SA and the opponent selects CA, the focal player gains
$d-q$. If the focal player selects CA and the opponent selects
SA, the focal player gains $1-d$. If both players select CA,
both players obtain $1/2-q$. Because $1-d>1/2>1/2-q>d-q$ and
$2\times (1/2)>(1-d)+(d-q)$, the single-shot crowdsourcing dilemma game played by
strategy 14 and uncond-CA is identified with the prisoner's dilemma.

The uncond-CA strategy is equivalent to the unconditional defector in the prisoner's dilemma. To describe the behavior of the player adopting strategy 14, we refer to such a player simply as strategy 14 here.
If strategy 14 realizes SA and the opponent realizes SA, corresponding to mutual cooperation, strategy 14 selects SA (i.e., cooperation) in the next round.
If strategy 14 realizes SA and the opponent realizes CA, strategy 14 is exploited by the opponent and switches to CA (i.e., defection) in the next round.
If strategy 14 realizes CA and the opponent realizes SA, strategy 14 exploits the opponent and continues to select CA.
If strategy 14 realizes CA and the opponent realizes CA, corresponding to mutual defection, strategy 14 switches to SA.
Therefore, strategy 14 is equivalent to the win-stay lose-shift strategy in the iterated prisoner's dilemma~\cite{Nowak-Sigmund-1993,Kraines-Kraines-1993}. Our results pertaining to the improved efficiency of strategy 14 relative to uncond-CA are consistent with the results obtained for the win-stay lose-shift strategy in the iterated prisoner's dilemma~\cite{Nowak-Sigmund-1993,Kraines-Kraines-1993}.
 
Intuitively, crowdsourcing and not attacking are both analogous to cooperation, and not crowdsourcing and attacking are analogous to defection.
However, the present model as well as the previous one \cite{Manuel} do not allow the association between crowdsourcing (not crowdsourcing) and cooperation (defection) because of the definition of the payoff. In both models,
the winning player gains a payoff equal to unity. Given that attacking does not occur, 
crowdsourcing increases the probability of winning owing to the enhanced productivity. However, whether the solution is made in-house or by crowdsourcing does not affect the payoff in any other way. For example, if the realized actions of both players are SN (i.e., not crowdsourcing and not attacking), each player gains an expected payoff equal to $1/2$. If the realized actions of the two players are CN (i.e., crowdsourcing and not attacking), the payoff remains the same.
However, in real situations, crowdsourcing is considered to improve the quality of the solution unless an attack occurs~\cite{von2008recaptcha,cooper2010predicting,barrington2012game}.
To examine non-iterated and iterated crowdsourcing dilemma games with this added component warrants future work. In this study, we confined ourselves to a simpler scenario, thus avoiding to introduce yet a new parameter.

\section*{Methods}

\subsection*{Calculation of average payoffs}

Throughout the present paper, 
we concentrate on the set of pure reactive strategies.
We calculate the average payoff of player 1 that adopts reactive strategy $n$ when the opponent player 2 adopts reactive strategy $m$, denoted by $\pi_{nm}$.

In each round, there are nine possible pairs of actions realized by the two players.
In other words, $(\theta_{1,t},\theta_{2,t})$ is either (CA,CA), (CA,CN), (CN,CA), (CN,CN), (C{\textasteriskcentered},SA), (C{\textasteriskcentered},SN), (SA,C{\textasteriskcentered}), (SN,C{\textasteriskcentered}), or (S{\textasteriskcentered},S{\textasteriskcentered}).
Given $(\theta_{1,t},\theta_{2,t})$, the actions that the two players intend to carry out in the next round are determined by $n$ and $m$.
Then, we calculate the probability with which each pair of actions $(\alpha_{1,t+1},\alpha_{2,t+1})$ is actually selected. This probability depends on
$\epsilon$.
Then, the two players play the game such that a pair of realized actions $(\theta_{1,t+1},\theta_{2,t+1})$ is uniquely determined from the pair of selected actions $(\alpha_{1,t+1},\alpha_{2,t+1})$, as shown in Table~\ref{submit_and_realize}.
By combining the stochastic mapping from
 $(\theta_{1,t},\theta_{2,t})$ to $(\alpha_{1,t+1},\alpha_{2,t+1})$ and the deterministic mapping from $(\alpha_{1,t+1},\alpha_{2,t+1})$ to $(\theta_{1,t+1},\theta_{2,t+1})$, we obtain the transition probability from $(\theta_{1,t},\theta_{2,t})$ to $(\theta_{1,t+1},\theta_{2,t+1})$.
Any $(\theta_{1,t+1},\theta_{2,t+1})$ is reached from
any $(\theta_{1,t},\theta_{2,t})$ with a positive probability
because of the error ($\epsilon>0$).
Therefore, the Markov chain on $(\theta_{1,t},\theta_{2,t})$ is ergodic and possesses a unique stationary distribution.

The average payoff of player 1 is given by
\begin{equation}
\pi_{nm}=\sum_{(\theta_{1,t},\theta_{2,t})}P^*(\theta_{1,t},\theta_{2,t})\pi_1(\theta_{1,t},\theta_{2,t}),\label{eq:ave_payoff}
\end{equation}
where $P^*(\theta_{1,t},\theta_{2,t})$ is the probability that $(\theta_{1,t},\theta_{2,t})$ is realized in the stationary state, and $\pi_1(\theta_{1,t},\theta_{2,t})$ is the expected payoff of player 1 under $(\theta_{1,t},\theta_{2,t})$.
Table~\ref{pay_matrix} shows the values of $\pi_1(\theta_{1,t},\theta_{2,t})$.
It should be noted that if both players crowdsource, player 1 attacks player 2, and player 2 does not attack player 1 (i.e., $(\theta_{1,t},\theta_{2,t})=(\text{CA},\text{CN})$), then player 1 wins if $p_1>p_2-d$, where $p_1$ and $p_2$ are productivity values of players 1 and 2, respectively.
This event occurs with probability $1-(1/2)(1-d)^2$.

\subsection*{Evolutionary stability and efficiency}

Strategy $n$ is an ESS if $\pi_{nn}>\pi_{mn}$ is satisfied or both $\pi_{nn}=\pi_{mn}$ and $\pi_{nm}>\pi_{mm}$ are satisfied for all $m\neq n$.
We enumerate all ESSs as follows using essentially the same exhaustive search method as that used for studying indirect reciprocity \cite{Ohtsuki-Iwasa-2004}.  

Consider strategy $n$.
For all strategies $m\neq n$, we check the following conditions.
If $\pi_{nn}$ and $\pi_{mn}$ are not the same function in terms of $d$, $q$, and $\epsilon$, we expand the difference with respect to $\epsilon$ as follows: 
\begin{equation}
  \pi_{nn}-\pi_{mn}=\sum_{k=0}^{\infty}{a_k(d,q)\epsilon^k}.\label{taylor}
\end{equation}
We denote the nonzero coefficient of the lowest order on the right-hand side of Eq. (\ref{taylor}) by $a_{\overline{k}}(d,q)$.
For infinitesimally small $\epsilon$, strategy $n$ is stable against invasion by strategy $m$ if $a_{\overline{k}}(d,q)>0$.
If $\pi_{nn}$ and $\pi_{mn}$ are the same function, then we compare $\pi_{nm}$ and $\pi_{mm}$ via the same procedure.
If $\pi_{nm}$ and $\pi_{mm}$ are the same function, strategy $n$ is not an ESS because it is neutrally stable against invasion by strategy $m$.
If $n$ is not invaded by any $m$, even neutrally, $n$ is an ESS.

We say that ESS $n$ is efficient if $\pi_{nn}\geq \pi_{mm}$ for all other ESSs 
$m (\neq n)$.
To check the efficiency of ESSs, we expand $\pi_{nn}-\pi_{mm}$ with respect to $\epsilon$ and look at the sign of the non-zero coefficient of the lowest order.

\subsection*{Calculation of the size of the attractive basin under replicator dynamics}

We denote the frequency of players adopting strategy $n$ by $x_n \in [0,1]$.
The replicator equation is given by
\begin{equation}
\frac{dx_n}{dt}=x_n(\pi_n-\bar{\pi}),
\end{equation}
where $\pi_n=\sum_m \pi_{nm}x_m$ and $\bar{\pi}=\sum_{n}\pi_nx_n$. 

Consider the case in which just two pure-strategy ESSs, denoted by $n=1$ and 2, exist.
Then, under the replicator dynamics composed of two strategies 1 and 2, the relative size of the attractive basin of strategy 1 is given by $(\pi_{11}-\pi_{21})/(\pi_{11}+\pi_{22}-\pi_{12}-\pi_{21})$.

Consider the case in which three pure-strategy ESSs, denoted by $n=1$, 2, and 3, exist.
Then, we assume a population composed of the three ESSs and determine the basin size of the ESSs by direct numerical integration of the replicator equation because analytical expressions are difficult to obtain.
We run the dynamics from initial conditions $(x_1,x_2,x_3)=(\ell_1\Delta, \ell_2\Delta, \ell_3\Delta)$, where $\Delta=1/200$, and $(\ell_1, \ell_2, \ell_3)$ is a set of integers that satisfy $0< \ell_1, \ell_2, \ell_3 < 200$ and $\ell_1+\ell_2+\ell_3=200$.
We count the number of initial conditions such that all players finally adopt strategy 1 and divide it by the total number of initial conditions.
The calculated fraction defines the relative size of the attractive basin of strategy 1.
Parallel definitions are applied to strategies 2 and 3.


\section*{Acknowledgements}
K.O.
acknowledges the support provided through CREST JST.
M.C. and A.A. are supported by the Australian Government as represented by DBCDE and ARC through the ICT Centre of Excellence program.
N.M. acknowledges the support provided through Grants-in-Aid for Scientific Research (No. 23681033) from MEXT, Japan, the Nakajima Foundation, CREST JST, and the Aihara Innovative Mathematical Modelling Project, the Japan Society for the Promotion of Science (JSPS) through the ``Funding Program for World-Leading Innovative R\&D on Science and Technology (FIRST Program),'' initiated by the Council for Science and Technology Policy (CSTP).

\section*{Author contributions}

M.C. and N.M. designed the research; K.O. contributed the computational results; K.O., M.C., A.A., and N.M. discussed the results; K.O., M.C., A.A., and N.M. wrote the paper.

\section*{Additional information}

\textbf{Competing financial interests:} The author declares no competing financial interests.

\clearpage

\clearpage

\clearpage

\clearpage

\clearpage

\begin{table}
  \caption{Relationship between the actions selected by the two players and the realized actions.}
  \label{submit_and_realize}
    \begin{center}
      \begin{tabular}{cccc}
        \toprule
        $\alpha_{1,t}$&$\alpha_{2,t}$&$\theta_{1,t}$&$\theta_{2,t}$\\
        \midrule
        CA&CA&CA&CA\\
        CA&CN&CA&CN\\
        CA&SA&C{\textasteriskcentered}&SA\\
        CA&SN&C{\textasteriskcentered}&SN\\

        CN&CA&CN&CA\\
        CN&CN&CN&CN\\
        CN&SA&C{\textasteriskcentered}&SA\\
        CN&SN&C{\textasteriskcentered}&SN\\

        SA&CA&SA&C{\textasteriskcentered}\\
        SA&CN&SA&C{\textasteriskcentered}\\
        SA&SA&S{\textasteriskcentered}&S{\textasteriskcentered}\\
        SA&SN&S{\textasteriskcentered}&S{\textasteriskcentered}\\

        SN&CA&SN&C{\textasteriskcentered}\\
        SN&CN&SN&C{\textasteriskcentered}\\
        SN&SA&S{\textasteriskcentered}&S{\textasteriskcentered}\\
        SN&SN&S{\textasteriskcentered}&S{\textasteriskcentered}\\
        \bottomrule
      \end{tabular}
    \end{center}
\end{table}

\newpage
\clearpage

\begin{table}
  \caption{The expected payoff $\pi_1(\theta_{1,t},\theta_{2,t})$ of player 1 when the realized actions of player 1 and 2 are $\theta_{1,t}$ and $\theta_{2,t}$, respectively.}
  \label{pay_matrix}
    \begin{center}
      \begin{tabular}{ccl}
        \toprule
        $\theta_{1,t}$&$\theta_{2,t}$&$\pi_1(\theta_{1,t},\theta_{2,t})$\\
        \midrule
        CA&CA&$1/2-q$\\
        CA&CN&$1-(1/2)(1-d)^2-q$\\
        CN&CA&$(1/2)(1-d)^2$\\
        CN&CN&$1/2$\\
        C{\textasteriskcentered}&SA&$1-d$\\
        C{\textasteriskcentered}&SN&$1$\\
        SA&C{\textasteriskcentered}&$d-q$\\
        SN&C{\textasteriskcentered}&$0$\\
        S{\textasteriskcentered}&S{\textasteriskcentered}&$1/2$\\
        \bottomrule
      \end{tabular}
    \end{center}
\end{table}

\newpage
\clearpage
\newcolumntype{x}{>{\centering\arraybackslash}p{9mm}}
\begin{table}
    \caption{ESSs. $\alpha_n(\theta)$ is the action that a player with strategy $n$ selects when the opponent's realized action was $\theta$ in the previous round.
      The regions for ESS and efficiency in the table indicate the parameter regions in which the strategy is an ESS and an efficient ESS, respectively.
      See Figure~\ref{region_plot} for the definition of the region labels.
      If an ESS is efficient nowhere, the entry for the region for efficiency remains blank.}
    \label{ess_def_region}
    \begin{center}
      \begin{tabular}{lxxxxxxxc}
        \toprule
        Strategy ($n$) &$\alpha_n(\text{CA})$&$\alpha_n(\text{CN})$&$\alpha_n(\text{C{\textasteriskcentered}})$&$\alpha_n(\text{SA})$&$\alpha_n(\text{SN})$&$\alpha_n(\text{S{\textasteriskcentered}})$&\shortstack{Region\\for ESS}&\shortstack{Region for\\efficiency}\\ \hline
        1 (uncond-CA)&CA&CA&CA&CA&CA&CA&(A)&\\
        2 (uncond-CN)&CN&CN&CN&CN&CN&CN&(B)&(B)\\
        3 (uncond-SA)&SA&SA&SA&SA&SA&SA&(C)&(C)\\

        4&CN&CN&CN&CN&CN&SA&(D)&(D)\\
        5&CN&CN&SA&SA&SA&CN&(D)&(D)\\
        6&CN&CN&SA&SA&SA&SA&(D)&(D)\\
        7&SA&SA&CN&CN&CN&CN&(D)&(D)\\
        8&SA&SA&CN&CN&CN&SA&(D)&(D)\\
        9&SA&SA&SA&SA&SA&CN&(D)&(D)\\

        10&CN&SA&SA&SA&SA&SA&(D)&\\
        11&CN&SA&CN&CN&CN&SA&(E)&\\

        12&CA&CA&CA&CA&CA&SA&(F)&(K)\\
        13&SA&SA&CA&CA&CA&CA&(G)&\\
        14&SA&SA&CA&CA&CA&SA&(H)&(L)\\
        15&SN&CA&CA&CA&CA&SN&(I)&\\
        16&SN&CN&CA&CA&CA&SN&(J)&\\
        \bottomrule
      \end{tabular}
    \end{center}
\end{table}

\newpage
\clearpage
\newcolumntype{y}{>{\centering\arraybackslash}p{8mm}}
\begin{table}
\caption{Stationary distribution of the selected actions of two players adopting an ESS, in the limit $\epsilon \to 0$. $P(\alpha)$ is the probability that both players select action $\alpha$. The probability that the two players select different actions tends to zero as $\epsilon\to 0$. The payoff when both players adopt the same strategy is also shown.}
    \label{tab:stationary}
    \begin{center}
      \begin{tabular}{lyyyyl}
        \toprule
        Strategy ($n$) &$P(\text{CA})$&$P(\text{CN})$&$P(\text{SA})$&$P(\text{SN})$& Payoff ($\pi_{nn})$\\
        \midrule
        1 (uncond-CA)&1&0&0&0&$(1/2-q)+2q\epsilon-q\epsilon^2$\\
        2 (uncond-CN)&0&1&0&0&$1/2-q\epsilon+q\epsilon^2$\\
        3 (uncond-SA)&0&0&1&0&$1/2-q\epsilon+q\epsilon^2$\\
        4&0&1&0&0&$1/2-q\epsilon+q\epsilon^2$\\
        5&0&1&0&0&$1/2-q\epsilon+q\epsilon^2$\\
        6&0&0&1&0&$1/2-q\epsilon+q\epsilon^2$\\
        7&0&1/2&1/2&0&$1/2-q\epsilon+q\epsilon^2$\\
        8&0&0&1&0&$1/2-q\epsilon+q\epsilon^2$\\
        9&0&1/2&1/2&0&$1/2-q\epsilon+q\epsilon^2$\\
        10&0&0&1&0&$1/2-q\epsilon+q\epsilon^2-2q\epsilon^3+\mathcal{O}(\epsilon^4)$\\
        11&0&0&1&0&$1/2-q\epsilon-q\epsilon^2+8q\epsilon^3+\mathcal{O}(\epsilon^4)$\\
        12&1&0&0&0&$(1/2-q)+(5/2)q\epsilon-(5/2)q\epsilon^2+q\epsilon^3+\mathcal{O}(\epsilon^4)$\\
        13&1/2&0&1/2&0&$(1/2)(1-q)+(3/2)q\epsilon^2-q\epsilon^3+\mathcal{O}(\epsilon^4)$\\
        14&0&0&1&0&$1/2-3q\epsilon+9q\epsilon^2-10q\epsilon^3+\mathcal{O}(\epsilon^4)$\\
        15&0&0&0&1&$1/2-2q\epsilon+20q\epsilon^3+\mathcal{O}(\epsilon^4)$\\
        16&0&0&0&1&$1/2-2q\epsilon+(1/4)q\epsilon^2+(399/16)q\epsilon^3+\mathcal{O}(\epsilon^4)$\\
        \bottomrule
      \end{tabular}
    \end{center}
\end{table}

\end{document}